\documentclass[12pt]{article}
\usepackage{bm}
\date{}                   

\everymath{\displaystyle}
\begin{document}           

\begin{center}

{\bf Quantum-Mechanical Description of Spin-1/2 Particles
and Nuclei Channeled in Bent Crystals}

{\bf A. J. Silenko}
\vskip 3mm

\emph{Joint Institute for Nuclear Research, Dubna, 141980 Russia \\ Research Institute for Nuclear Problems, Belarusian State University, Minsk, 220030, Belarus\\
E-mail: alsilenko@mail.ru}

\end{center}
\vskip 3mm

General quantum-mechanical description of relativistic particles and nuclei with spin 1/2 channeled in bent crystals is performed with the use of the cylindrical coordinate system. The previously derived Dirac equation in this system is added by terms characterizing anomalous magnetic and electric dipole moments. A transformation to the Foldy-Wouthuysen re\-pre\-sen\-ta\-tion, a derivation of the quantum-mechanical equations of motion for particles and their spins, and a determination of classical limit of these equations are fulfilled in the general case. A physical nature of main peculiarities of description of particles and nuclei in the cylindrical coordinate system is ascertained.
\vskip 3mm


\section{Introduction}

The strict quantum-mechanical description of relativistic spin-1/2 particles and nuclei channeled in
bent crystals is an important problem \cite{Gang}. Taking spin
effects into account is very important in such a
description. It is known that, during planar channeling
in bent crystals, the particle and nucleus spins are
rotated by a rather large angle. This effect was first
found in the papers of Baryshevsky \cite{BarJAPL,BarJPhys}, where he
also proposed its use to determine the magnetic
moments of shortliving particles. The simple dependence between the rotation angles of particles and
their spins in bent crystals was found by Lyuboshits \cite{Lubo}. In such crystals, the centrifugal force acting on particles or nuclei moving along bent trajectories is compensated by the Coulomb force, which leads to the
appearance of a rather strong electric field rotating the
spin. The effect of spin rotation was observed experimentally in \cite{CAB,KSCC}.

Although particle and nucleus channeling in many
cases can be adequately described by methods of classical theory, a thorough quantum-mechanical analysis
of the problem is also necessary. So, the discreteness of
the energy spectrum is very often important for relativistic positrons (electrons). To determine a given spectrum correctly, it is necessary to adequately take into
account spin effects for relativistic particles. As is well
known, in the Dirac equation, as in the Dirac-Pauli
equation, which takes the anomalous magnetic
moment (AMM) into account, Dirac matrices determine the spin projections on the axes of the Cartesian
system of coordinates. It is convenient to use such a system of coordinates only for channeling in unbent
crystals, and the choice of the cylindrical system of
coordinates is natural for bent crystals (if the radius of
curvature is approximately constant).

The author of \cite{JETP1995} presented the quantum-mechanical description of spin 1/2 particles (nuclei) for planar
channeling in unbent and bent crystals. Special attention was focused on determination of the spin dynamics. In this paper, the author solved the Dirac equation
(supplemented with terms describing the AMM) in
the Foldy-Wouthhuysen (FW) representation, constructed the operator equation of spin motion, and
calculated the average value of its precession frequency. The results obtained using the quantum-mechanical description agreed entirely with the corresponding classical results. The author of \cite{JETP1995} used the
Cartesian system of coordinates rather than the cylindrical one and took the presence of crystal bending
into account formally by including an additional
potential energy determining the correction for the
centrifugal force into the Hamilton operator in the
FW representation.

Naturally, such an approach is not rigorous,
although it leads to reasonable results. In the present paper,
the Dirac equation in the cylindrical system of coordinates derived in \cite{Schluter} is used as the initial one. We supplement it with terms describing the AMM and the
electric dipole moment (EDM) and perform transformation into the FW representation by a method developed in \cite{JMP}. We use the obtained Hamilton operator in
this representation to derive general equations describing the motion of particles and nuclei and the
spin evolution.

We let the respective Greek and Roman letters
$\alpha,\mu,\nu,\ldots=0,1,2,3$ and $i,j,k,\ldots=1,2,3$ denote the
world and space indices in four-dimensional spacetime. Using the apparatus of the theory of general relativity, tetrad indices are denoted by the initial letters
of the Roman alphabet $a,b,c,\ldots = 0,1,2,3$. The time and
space tetrad indices are singled out by hats. The signature has the form $(+---)$. We here use the system of
units $\hbar=1,~c=1$. In some cases, to make the presentation clearer, we include Planck’s constant in the
corresponding formulas. The notations $[\dots,\dots]$ и $\{\dots,\dots\}$ determine the commutators and anticommutators, respectively.

\section{Dirac-Pauli equations in the cylindrical system of coordinates}

The standard Dirac–Pauli equation (in the Cartesian system of coordinates) has the form
\begin{equation} \biggl[\gamma^\mu\pi_\mu-m+\frac{\mu'}{2}\sigma^{\mu\nu}F_{\mu\nu}\biggr] \Psi=0,\label{gDPeq}
\end{equation}
where $\gamma^\mu$ and $\sigma^{\mu\nu}=i(\gamma^\mu\gamma^\nu-\gamma^\nu\gamma^\mu)/2$ are the Dirac
matrices, $F_{\mu\nu}=(\bm E,\bm B)$ is the electromagnetic field tensor, $\mu'$ is the AMM, and $\pi_\mu=iD_\mu=i\hbar(\partial/\partial x^\mu)-eA_\mu$. Here $\bm E,\bm B$ and $A^\mu$ are the electric strength, the magnetic induction, and the four-potential of the electromagnetic field. In \cite{RPJ}, this equation was supplemented with a term describing the EDM $d$:
\begin{equation} \biggl[\gamma^\mu\pi_\mu-m+\frac{\mu'}{2}\sigma^{\mu\nu}F_{\mu\nu}
+\frac{d}{2}\sigma^{\mu\nu}G_{\mu\nu}\biggr] \Psi=0,\label{gDPEqDM}
\end{equation}
where $G_{\mu\nu}=(-\bm B,\bm E)$ is the tensor dual to $F_{\mu\nu}$. 

In principle, it is possible to pass to cylindrical
coordinates, not changing the definition of the Dirac
matrices and using the relations
$$\begin{array}{c}
\gamma^{\rho}=\bm\gamma\cdot\bm e_\rho=\gamma^x\cos{\phi}+\gamma^y\sin{\phi},~~~ 
\gamma^{\phi}=\bm\gamma\cdot\bm e_\phi=-\gamma^x\sin{\phi}+\gamma^y\cos{\phi}.
\end{array}$$
Naturally, such a way is not convenient, and the subsequent transformations are accompanied by very cumbersome calculations. Of course, all required calculations can be carried out in the Cartesian system of
coordinates. However, to specify an external field, if
the symmetry of the problem is taken into account, it
is more convenient to use the cylindrical coordinates
specifically. A convenient form of the Dirac equation
in the cylindrical system of coordinates was found in \cite{Schluter}, the authors of which used the fundamental Pauli
theorem \cite{PauliFT} determining the relationship between
different sets of Dirac matrices satisfying the required
commutation and anticommutation relations. The
Dirac equation derived in \cite{Schluter} (for $\mu'=0$) has a very simple form and formally coincides with the initial
equation:
\begin{equation} (\gamma^\mu\pi_\mu-m) \Psi=0.\label{Dcyl}
\end{equation}
The matrices $\gamma^\mu$ are ordinary Dirac matrices. However,
here, the indices 1, 2, and 3 correspond to the cylindrical coordinates $\rho,\phi$, and $z$, and
\begin{equation} (\pi_1,\pi_2,\pi_3)=\left(i\hbar\frac{\partial}{\partial\rho}-eA_\rho,\,i\hbar\frac1\rho\frac{\partial}{\partial\phi}-eA_\phi,\,i\hbar\frac{\partial}{\partial z}-eA_z\right).\label{pi}
\end{equation} The contravariant vector potential $\bm A$, whose components are $-A_i$ is usually used.

It follows from Eq. (\ref{Dcyl}) that the operator $\pi_i$ contains
the nabla operator in the cylindrical system of coordinates. The Dirac equation in the spherical system of
coordinates has similar properties \cite{Schluter}.
The result
obtained in \cite{Schluter} is completely natural and shows that
the transformation into cylindrical and spherical
coordinates which retains the form of the Dirac matrices $\gamma^\mu$does not violate the covariance of the Dirac
equation.

We use this fact to include terms proportional to the
AMM and the EDM into the equation. Such inclusion
will be substantiated additionally in the next section by
comparison with the results obtained within the
framework of the theory of general relativity.

The covariance of Eq. (\ref{Dcyl}) is not violated if its generalization is written in a form (\ref{gDPEqDM}), where the matrices $\sigma^{\mu\nu}$ have the usual form and the indices $\mu$ and $\nu$ correspond to the cylindrical system of coordinates. The
components of the tensors $F_{\mu\nu}$ and $G_{\mu\nu}$ are also
determined in this system of coordinates.

It is convenient to multiply the obtained equation
by the matrix $\gamma^0$ and represent it into the Hamiltonian form
\begin{equation}\begin{array}{c} i\hbar\frac{\partial\Psi}{\partial t}={\cal H}\Psi, ~~~ {\cal H}=\bm{\alpha}\cdot\bm{\pi}+\beta
m+e\Phi+\mu'(-\bm {\Pi}\cdot \bm{B}+i\bm{\gamma}\cdot\bm{
E})-d(\bm {\Pi}\cdot \bm{E}+i\bm{\gamma}\cdot\bm{
B}),\\ \bm{\pi}=-i\hbar\nabla-e\bm{A}, \end{array}\label{eq19DiP}
\end{equation}
where $\Phi\equiv A_0$ is the scalar potential and $\bm{\pi}=-(\pi_\rho,\pi_\phi,\pi_z)$ is the kinetic momentum operator in the
cylindrical system of coordinates. Unlike the Cartesian system of coordinates, the different components
of the operators $\bm{\pi}$ and $\nabla$ do not commute with each
other in the general case.

Equation (\ref{eq19DiP}) is initial for the next transition to the
FW representation. Nevertheless, the obtained result
requires additional substantiation because the terms
describing the AMM and EDM were included in
(\ref{Dcyl}) obtained as a result of the transformation, rather than in the standard Dirac equation. Such substantiation can be obtained within the framework of
gravitation theory. Using gravitation theory makes it possible to relatively simply understand the physical
meaning of the difference between the dynamics of
particles in the Cartesian and cylindrical systems of
coordinates.

\section{Deriving the Dirac-Pauli
Hamiltonian for particles in the cylindrical system of coordinates using the methods
of gravity}

The correctness of the above generalization of
Eq. (\ref{Dcyl}), which makes it possible to take into account
the possible presence of the AMM and the EDM of
the particle, can be confirmed by analyzing the covariant Dirac equation in gravitation theory. This equation describes the electromagnetic interaction of the
Dirac particle in Riemannian spacetime and has the
form \cite{HN,BlagojevicHehl}
\begin{equation}
(i\hbar\gamma^a D_a - m)\psi=0,\qquad a=0,1,2,3.
\label{Dirac0}
\end{equation}
In Eq. (\ref{Dirac0}) we have
\begin{equation}\begin{array}{c} D_a = e_a^\mu\left(
\partial_\mu + \frac{ie}{\hbar}A_\mu\right)+{\frac i4}\sigma^{bc}\Gamma_{bca},
\end{array}\label{eqin2}\end{equation}
where $\Gamma_{abc}=-\Gamma_{bac}$ are the Ricci rotation coefficients \cite{LL2}) having the following form:
\begin{equation}\begin{array}{c}
\Gamma_{abc}=\frac 12\left(C_{abc}-C_{bca}-C_{cab}\right),~~~
C_{abc}=e_a^\mu e_b^\nu (\partial_\mu e_{c\nu}-\partial_\nu
e_{c\mu}).
\end{array}\label{eqin7}\end{equation} Here $e_a^\mu$ are tetrad coefficients determining the tetrad
components of the covariant derivative ($D_a =e_a^\mu D_\mu$) and other four-vectors.

General equation (\ref{Dirac0}) can be used for cylindrical
and other curvilinear coordinates. The transition from $x,y,z$ to $\rho,\phi,z$ in the expression for the
squared infinitesimal interval $ds^2$ gives the following
form for the metric tensor:
\begin{equation}g_{\mu\nu}=\left(\begin{array}{cccc}1&0&0&0\\0&-1&0&0\\0&0&-\rho^2&0\\0&0&0&-1\end{array}\right).
\label{metrite}\end{equation} Metric tensor (\ref{metrite}) describes a flat spacetime.

In this case, the nonzero Ricci rotation coefficients are
\begin{equation}\Gamma_{212}=-\Gamma_{122}=\frac1\rho.
\label{Riccinz}\end{equation}
The covariant Dirac equation obtained by substituting (\ref{Riccinz}) into (\ref{Dirac0}) and (\ref{eqin2}) does not coincide with Eq. (\ref{Dcyl}). However, we note that to find the Hermitian Hamiltonian, it is necessary to perform the nonunitary transformation of the wave function of the initial covariant
Dirac equation (\ref{Dirac0}) which has the form \cite{ostrong}
\begin{equation}
\Psi = \left(\sqrt{-g}e_{\widehat{0}}^0\right)^{1/2}\psi=\sqrt{\rho}\psi,\label{newwftn}
\end{equation} where $g$ is the determinant of the metric tensor. To
obtain the Hermitian Hamiltonian from Eq. (\ref{Dcyl}), such
a transformation is not required.

The natural generalization of Eq. (\ref{Dirac0}) making it
possible to describe the AMM and the EDM and
retaining the equation covariance has the form
\begin{equation}
\left(i\hbar\gamma^a D_a - m + {\frac{\mu'}{2}}
\sigma^{ab}F_{ab} + {\frac {d}{2}}\sigma^{ab}G_{ab}\right)\Psi=0.\label{DiracPauli}
\end{equation}
Here, $F_{ab}=e_a^\mu e_b^\nu F_{\mu\nu},~G_{ab}=e_a^\mu e_b^\nu G_{\mu\nu},$ and $\sigma^{ab}=i(\gamma^a\gamma^b-\gamma^b\gamma^a)/2 $.

In the general case, Eq. (\ref{Dirac0}) was transformed into
the Hamilton form in \cite{ostrong}, and the authors of \cite{OSTgrav} transformed the obtained Hamiltonian into the FW
representation also in the general case. For metric (\ref{metrite}), using the method for finding the Hamilton form of
Eq. (\ref{DiracPauli}) proposed in \cite{ostrong} leads to an expression for the
Hermitian Hamiltonian which coincides with (\ref{eq19DiP}). Thus, the correctness of Eq. (\ref{eq19DiP}) is completely confirmed strictly by the methods of quantum mechanics
of Dirac particles in gravitational fields.

Gravitation theory also enables us to determine the
distinguishing features of particle dynamics in the
cylindrical system of coordinates. The particle velocity
in this system has the form $\bm v=v_\rho \bm e_\rho +v_\phi \bm e_\phi +v_z \bm e_z$. Naturally, its constancy ($\bm v=const$) means that a force,
keeping the particle in a circular orbit, acts on it. This
force can be determined using the formalism of gravitoelectromagnetic fields which makes it possible to
describe relativistic particles in arbitrarily strong gravitational fields. This formalism was proposed by
Pomeranskii and Khriplovich \cite{PK} and underwent further development in \cite{ostrong,OSTgrav,Warszawa,OST}. It is convenient
to introduce gravitoelectromagnetic fields to determine the equations of motion described in local
Lorentz (tetrad) reference systems:
\begin{equation}
\begin{array}{c} \frac{d\bm s}{dt}=\bm{\Omega}\times \bm s, ~~~ \bm{\Omega}=\frac{1}{u^0}\left(-\bm{\mathcal{B}}+
\frac{\widehat{\bm u}\times\bm{\mathcal{E}}}
{u^{\widehat{0}}+1}\right),
\end{array} \label{omgem}
\end{equation}
\begin{equation}
\begin{array}{c} \frac{d\widehat{\bm{u}}}{dt}=\frac{u^{\widehat{0}}}{u^0}\left(\bm{\mathcal{E}}+
\frac{\widehat{{\bm
u}}\times\bm{\mathcal{B}}}{u^{\widehat{0}}}\right),  ~~~ \frac{d{u}^{\widehat{0}}}{dt}=
\frac{\bm{\mathcal{E}}
\cdot\widehat{{\bm
u}}}{u^0}.
\end{array} \label{force} \end{equation}
These equations are analogous to equations for a particle with a Dirac magnetic moment ($g=2$) in an electromagnetic field. The spin $\bm s$, the gravitoelectric field $\bm{\mathcal{E}}$, and the gravitomagnetic field $\bm{\mathcal{B}}$ are determined precisely in these local Lorentz reference systems; i.e.,
their components are of the tetrad type. The general
equation for the fields has the form
\begin{equation}
\begin{array}{c} \mathcal{E}_{\widehat{i}}=\Gamma_{0ic}u^{c},~~~
\mathcal{B}_{\widehat{i}}=-\frac{1}{2}e_{ikl}\Gamma_{klc}u^{c},
\end{array} \label{expl}
\end{equation}  where $e_{ikl}$ is the antisymmetric tensor with spatial
components.

The tetrad and world directions coincide in terms
of the problem under consideration, and the fields are
determined by the expressions
\begin{equation}
\begin{array}{c} \bm{\mathcal{E}}=0,~~~
\mathcal{B}_{\widehat{\rho}}=\mathcal{B}_{\widehat{\phi}}=0,~~~\mathcal{B}_{\widehat{z}}=\frac{u^{\widehat{\phi}}}{\rho}.
\end{array} \label{expcylf}
\end{equation} Because $e^\phi_{\widehat{\phi}}=1/\rho$, we have $u^\phi=u^{\widehat{\phi}}/\rho$.

Equations (\ref{force}) and (\ref{expcylf}) show that the force determined by the gravitomagnetic field acting in the
cylindrical system of coordinates is an analogue of
the Lorentz force. Its appearance is a consequence of
the fact that, if the azimuthal angle of the particle
changes by $d\phi$, the horizontal axes of the cylindrical
and Cartesian systems of coordinates rotate by the
same angle with respect to each other; i.e., the cylindrical system of coordinates rotates with an instantaneous angular velocity $-d\phi/dt=-v_\phi/\rho$ with respect to the Cartesian one \cite{RPJSTAB}.

For the detailed quantum-mechanical description
of spin 1/2 particles and nuclei, it is convenient to use
the FW representation, the transition to which is performed in the following section.

\section{Dirac-Pauli Hamiltonian in the Foldy-Wouthuysen representation
for particles in the cylindrical system of coordinates}

The method of FW transformation for relativistic
particles in arbitrary external fields developed in \cite{JMP,PRA,JINRLClL,TMPFW}, is based on the following representation of
the initial Hamiltonian:
\begin{equation} {\cal H}_D=\beta m+{\cal E}+{\cal
O}, ~~~  \beta{\cal E}={\cal E}\beta, ~~~  \beta{\cal O}=-{\cal O}\beta, \label{eq3Dirac} \end{equation}
where $\beta\equiv\gamma^0$. In this equation, the initial Hamiltonian
is divided into even $({\cal E})$ and odd $({\cal O})$ terms, which are
diagonal and off-diagonal in two spinors, respectively.
In Eq. (\ref{eq3Dirac}), we have
\begin{equation} {\cal E}=e\Phi-\mu'\bm
{\Pi}\cdot \bm{B}-d\bm {\Pi}\cdot \bm{E}, ~~~{\cal
O}=\bm{\alpha}\cdot\bm{\pi}+i\mu'\bm{\gamma}\cdot\bm{
E}-id\bm{\gamma}\cdot\bm{B}. \label{DireqEDM} \end{equation}

The FW transformation proved itself to be the best
method for finding the quasiclassical approximation
and the classical limit of relativistic quantum mechanics in the case of the single-particle approach. In this
representation, the Hamiltonian is diagonal in two
spinors (block-diagonal), the probability interpretation of the wave function is restored, and the operators
have the same form as in nonrelativistic quantum
mechanics. The transition to the FW representation is
widely used for all fundamental interactions.

If the terms that are bilinear in external fields $\bm E$ and $\bm B$ are disregarded, the result of the transformation can
be written in the form
\begin{equation}\begin{array}{c}
{\cal H}_{FW}=\beta\epsilon+ {\cal E}-\frac 18\left\{\frac{1}
{\epsilon(\epsilon+m)},[{\cal O},[{\cal O},{\cal
F}]]\right\},~~~
\epsilon=\sqrt{m^2+{\cal O}^2}. \end{array} \label{eqfin}
\end{equation}
This Hamiltonian in the FW representation contains
exactly determined terms of the first and second orders
with respect to $\hbar/S_0$, where $S_0$ is the value of the action
dimensionality \cite{TMPFW}. Terms of second and higher order
with respect to $\hbar/S_0$ are also determined exactly in the
case where they appear as a result of calculating the
given Hamiltonian. In particular, this is related to the
Darwin (contact) interaction.

For the problem under consideration, the commutator of the operators $\pi_i$ and $\pi_j$ is equal to the sum of
the two terms:
$$[\pi_i,\pi_j]=-\hbar^2[\nabla_i,\nabla_j]-ie\hbar(\nabla_iA_j-\nabla_jA_i)=-\hbar^2[\nabla_i,\nabla_j]+ie\hbar e_{ijk}B^k.$$
The first term is equal to zero if Cartesian coordinates
are used. For cylindrical coordinates, the commutator
$$[\nabla_\rho,\nabla_\phi]=-\frac{1}{\rho^2}\frac{\partial}{\partial\phi}=\frac{p_\phi}{i\hbar\rho}$$ has a nonzero value.

As a result, the operator ${\cal O}^2$ is defined by the following exact expression:
\begin{equation}\begin{array}{c}
{\cal O}^2=\bm{\pi}^2-e\hbar\bm\Sigma\cdot\bm B-\hbar\Sigma_z\frac{p_\phi}{\rho}+\beta
\left(\bm\Sigma\cdot[\bm\pi\times\bm G]-\bm\Sigma\cdot[\bm
G\times\bm\pi]-\hbar\nabla\cdot\bm G\right)+\bm
G^2, \end{array} \label{osquare}
\end{equation} where $\bm G=\mu'\bm{E}-d\bm{B}$.

When calculating the general expression for the
Hamiltonian in the FW representation, it is possible to
disregard terms of second and higher degrees with
respect to external fields and terms of third and higher
degrees with respect to the Planck constant. With this
accuracy, the Hamiltonian is determined by the equations
\begin{equation} {\cal H}_{FW}={\cal H}_{FW}^{(0)}+{\cal H}_{FW}^{(MDM)}+{\cal
H}_{FW}^{(EDM)}, \label{FWHamEDM} \end{equation}
\begin{equation}\begin{array}{c} {\cal H}_{FW}^{(0)}=\beta\epsilon'+e\Phi-\frac\hbar4\Pi_z\left\{\frac{1}{\epsilon'},\frac{p_\phi}{\rho}\right\},~~~  \epsilon'=\sqrt{m^2+\bm\pi^2},
\end{array} \label{Hzernew} \end{equation}
\begin{equation}\begin{array}{c} {\cal H}_{FW}^{(MDM)}=\frac
   14\left\{\left(\frac{\mu_0m}{\epsilon'
   +m}+\mu'\right)\frac{1}{\epsilon'},\biggl(\bm\Sigma\!\cdot\![\bm\pi\!
\times\!\bm E]-\bm\Sigma\!\cdot\![\bm E\!\times\!\bm\pi]-\hbar\nabla\!
\cdot\!\bm E\biggr)\right\}\\ -\frac
12\left\{\left(\frac{\mu_0m}{\epsilon'}
+\mu'\right), \bm\Pi\!\cdot\!\bm B\right\}\\
+\beta\frac{\mu'}{4}\left\{\frac{1}{\epsilon'(\epsilon'+m)},
\biggl[(\bm{B}\!\cdot\!\bm\pi)(\bm{\Sigma}\!\cdot\!\bm\pi)+ (\bm{\Sigma}
\!\cdot\!\bm\pi)(\bm\pi\!\cdot\!\bm{B})+2\pi\hbar(\bm\pi\!\cdot\!\bm j+
\bm j\!\cdot\! \bm\pi)\biggr]\right\},
\end{array} \label{eq33new} \end{equation}
\begin{equation}
\begin{array}{c}
{\cal H}_{FW}^{(EDM)}=-d\bm\Pi\!\cdot\!\bm E
+\frac{d}{4}\left\{\frac{1}{\epsilon'(\epsilon'+m)},
\biggl[(\bm{E}\!\cdot\!\bm\pi)(\bm{\Pi}\!\cdot\!\bm\pi)+ (\bm{\Pi}
\!\cdot\!\bm\pi)(\bm\pi\!\cdot\!\bm{E})\biggr]\right\} \\-\frac
d4\left\{\frac{1}{\epsilon'},\biggl(\bm\Sigma\!\cdot\![\bm\pi\!
\times\!\bm B]-\bm\Sigma\!\cdot\![\bm
B\!\times\!\bm\pi]\biggr)\right\},
\end{array} \label{EDMeq12} \end{equation}
where $\mu_0=e\hbar/(2m)$ is the Dirac magnetic moment
and $\bm j$ is the density of the external current satisfying the
Maxwell equation
$$\bm j=\frac{1}{4\pi}\left(\nabla\times\bm B-\frac{\partial \bm E}
{\partial t}\right).$$
The quantities ${\cal H}_{FW}^{(MDM)}$ and ${\cal H}_{FW}^{(EDM)}$ define the contributions of the magnetic and electric dipole
moments (the MDM and the EDM), respectively. In
accordance with the Maxwell equations
$$\nabla\cdot\bm B=0, ~~~ \nabla\times\bm E=-\frac{\partial \bm B}
{\partial t},$$ the EDM does not contribute to contact interactions
with external charges and currents.

Equations (\ref{FWHamEDM})--(\ref{EDMeq12}) give a general solution of the
quantum-mechanical description of the electromagnetic interaction of the Dirac particle in the cylindrical system of coordinates. Comparison with the
results in \cite{RPJ} shows that the operators ${\cal H}_{FW}^{(MDM)}$ and ${\cal H}_{FW}^{(EDM)}$ have the same form as in the Cartesian system
of coordinates, and a new spin-dependent term
appears in ${\cal H}_{FW}^{(0)}$.

\section{General equations of particle dynamics} 

Deriving quantum-mechanical equations of particle and nucleus dynamics in the cylindrical system of
coordinates has important specific features related to
the noncommutation of the operators $\nabla_\rho$ and $\nabla_\phi$. The
form of the dynamic equations is universal:
\begin{equation}\begin{array}{c}
\frac{d\bm\pi}{dt}=\frac{i}{\hbar}[{\cal
H}_{FW},\bm\pi] + \frac{\partial\bm\pi}{\partial
t}=\frac{i}{\hbar}[{\cal H}_{FW},\bm\pi]-e\frac{\partial\bm
A}{\partial t}, \label{eqme} \end{array} \end{equation} \begin{equation}\begin{array}{c}
\frac{d\bm\Pi}{dt}=\frac{i}{\hbar}[{\cal
H}_{FW},\bm\Pi]= 
\bm\Omega\times\bm\Pi,
\label{eqmspin} \end{array} \end{equation} where $\bm\Omega$ is the operator of the angular velocity of the
spin precession. However, the above noncommutation
leads to the appearance of new terms in dynamic
equations compared to the corresponding equations in
the Cartesian system of coordinates.

As is known, the total force acting on the particle
with spin has a spin-dependent component. This component is usually called the Stern-Gerlach force, and,
for a moving particle, it depends on the magnetic and
electric fields. Although the Stern-Gerlach force
leads to the important effect of division of the beam
into two beams with different polarizations, as a rule, it is small for charged particles as compared with the
Lorentz force $e(\bm E+\bm v\times\bm B)$. The author of \cite{JETP1995} showed
that, in the case of planar channeling, the influence of
spin-dependent terms in the Hamiltonian in the FW
representation on the particle trajectory is small and
was not observed experimentally. Therefore, we do not
take the spin-dependent force into account when
determining the dynamics of the kinetic momentum
of a particle.

In the approximation used, the equation of motion
for kinetic momentum (\ref{eqme}) takes the form
\begin{equation}\begin{array}{c}
\frac{d\bm\pi}{dt}=e\bm E+\beta\frac{e}{4}\left\{\frac{1}{\epsilon'},\left(\bm\pi\times\bm B-\bm B\times\bm\pi\right)\right\}+\bm{\mathcal{F}},\\
\bm{\mathcal{F}}=\Biggl(\frac{\beta}{2}\biggl\{\frac{1}{\epsilon'},\frac{\pi_\phi p_\phi}{\rho}\biggr\},-\frac{\beta}{4}\biggl\{\frac{1}{\epsilon'},\Bigl\{\pi_\rho,\frac{p_\phi}{\rho}\Bigr\}\biggr\}, 0 \Biggr). \label{eqmff} \end{array} \end{equation}
The first two terms in the obtained equation give
the operator expression for the Lorentz force, and the
third term describes the additional force acting in the
cylindrical system of coordinates. It can be written in
a more compact vector form. We introduce the operator $\bm O=O\bm e_z=(p_\phi/\rho)\bm e_z$, characterizing the orbital
motion of the particle about the $z$ axis. In this case,
\begin{equation}\begin{array}{c}
\bm{\mathcal{F}}=\frac{\beta}{4}\biggl\{\frac{1}{\epsilon'},\Bigl(\bm\pi\times\bm O-\bm O\times\bm\pi\Bigr)\biggr\}. \label{omegffi} \end{array} \end{equation}

The author of \cite{JINRLClL} showed that, if the conditions of
the quasiclassical approximation (the de Broglie wavelength is less than the characteristic size of the region
of external field nonuniformity) are satisfied, using
the FW representation it becomes possible to reduce
determination of the classical limit of equations of relativistic quantum mechanics to replacing operators in
the Hamiltonian and the quantum-mechanical equations of motion with the corresponding classical quantities. In this case, it is possible to disregard the noncommutation of the operators in the quantum-mechanical expressions. In the case under consideration,
$\bm O\rightarrow\epsilon'\bm\omega=m\gamma\bm\omega$, where $\bm\omega=\omega\bm e_z=(v_\phi/\rho)\bm e_z$ is
the instantaneous angular velocity of the orbital particle motion about the $z$ axis and $\gamma$ is the Lorentz factor.
The classical limit of the expression for the additional
force has the form
\begin{equation}\begin{array}{c}
\bm{\mathcal{F}}\rightarrow\,-\bm\omega\times\bm\pi. \label{cllim} \end{array} \end{equation}
It is easy to see that, for particle motion along a circular arc, this equation reproduces the centrifugal force $\pi_\phi v_\phi/\rho$, and the force $\bm{\mathcal{F}}$ changes the instantaneous
angular velocity of the orbital particle motion by the
quantity $-\bm\omega$.

Expression (\ref{cllim}) is completely analogous to the corresponding expression for the force acting in a rotating
reference system (Eq. (3.37) in Ref. \cite{OSTgrav}).\footnote{In Ref. \cite{OSTgrav}, as in Eqs. (\ref{eqmff}) and (\ref{cllim}), the force was defined as the time
derivative of the covariant momentum operator. Coriolis factor 2
appears in the corresponding equations for the acceleration and
the spatial component of the contravariant four-velocity \cite{ostrong,PRD2}.} The analogy between the cylindrical system of coordinates and the
rotating reference system was demonstrated previously in \cite{RPJSTAB} within the framework of the classical
approach.

However, it is important that the rotating reference
system can be used only within the framework of the
one-particle description, while the cylindrical system
of coordinates can also be used to describe a beam of
particles or nuclei with various energies. The presence
of such a possibility is an important advantage of the
cylindrical system of coordinates.

The force $\bm{\mathcal{F}}$ is similar to forces appearing in noninertial reference systems. However, the latter are real
forces, which, in particular, can be measured by a
dynamometer, while the force $\bm{\mathcal{F}}$ is fictitious. Its presence does not affect the dynamometer readings,
although it affects the particle and nucleus motion in
the cylindrical system of coordinates. An interesting
effect of the mutual influence of particle motion in the
directions $\bm e_\rho$ and $\bm e_\phi$ follows from Eqs. (\ref{eqmff}) and (\ref{cllim}). In
particular, oscillator motion in the radial direction
causes the appearance of an oscillating term in the
expression for $p_\phi$ and vice versa. This effect has kinematic nature. It is due to the rotation of the axes $\bm e_\rho$ and $\bm e_\phi$ (about the Cartesian system of coordinates) as the
azimuth $\phi$ changes, and it disappears at $p_\phi=0$ as follows from the definition of $\bm O$.

Spin motion in the
cylindrical system of coordinates has a simpler character than the dynamics of the kinetic momentum. The
angular velocity operator of the spin precession, which
is easily determined using Eq. (\ref{eqmspin}), has the form
\begin{equation} \bm\Omega=\bm\Omega^{(0)}+\bm\Omega^{(MDM)}+\bm\Omega^{(EDM)}, \label{FWHamOme} \end{equation} where
\begin{equation}\begin{array}{c} \bm\Omega^{(0)}=-\frac\beta2 \left\{\frac{1}{\epsilon '},\bm O\right\}, \end{array} \label{Omega} \end{equation}
\begin{equation}
\begin{array}{c} \bm\Omega^{(MDM)}=\frac{1}{2\hbar}\left\{\left(\frac{\mu_0m} {\epsilon
'+m}+\mu'\right)\frac{1}{\epsilon '},(\bm \pi\times\bm E-\bm
E\times\bm \pi)\right\}- \frac\beta\hbar\left\{\left(
\frac{\mu_0m}{\epsilon '}+\mu'\right),\bm
B\right\}\\+ \beta\frac{\mu'}{2\hbar}\left\{\frac{1}{\epsilon '(\epsilon
'+m)},\biggl((\bm B\cdot\bm \pi)\bm \pi+ \bm \pi(\bm \pi\cdot\bm
B)\biggr)\right\},
\end{array} \label{eq36JMP} \end{equation}
\begin{equation}
\begin{array}{c}
\bm\Omega^{(EDM)}=-\beta\frac{2d}{\hbar}\bm
E+\frac{d}{2\hbar}\left\{\frac{1}{\epsilon '(\epsilon
'+m)},\biggl((\bm
E\cdot\bm \pi) \bm \pi +\bm \pi(\bm \pi\cdot\bm E)\biggr)\right\}\\
-\frac{d}{2\hbar}\left\{\frac{1}{\epsilon
'},\left(\bm \pi\times\bm B-\bm B\times\bm \pi\right)\right\}.
\end{array} \label{EDMeq14} \end{equation}
For a particle with a positive total energy in the FW
representation, the lower spinor is zero. Therefore,
passing to the classical limit eliminates the $\beta$ matrices
(and also the anticommutators).

It is easy to see that $\bm\Omega^{(0)}\rightarrow-\bm\omega$ in the classical
limit. Thus, the passage from the Cartesian to the
cylindrical coordinates changes the instantaneous
angular velocities of orbital particle motion and the precession of its spin by the same quantity, so that the
difference between them remains the same. The forms
of the quantities $\bm\Omega^{(MDM)}$ and $\bm\Omega^{(EDM)}$ in the Cartesian \cite{RPJ}
and cylindrical systems of coordinates coincide.

\section{Motion of particle and nucleus spins at planar channeling in bent crystals}

As was mentioned above, the quantum-mechanical
description of spin 1/2 particles and nuclei during planar channeling in unbent and bent crystals was presented previously in \cite{JETP1995}, the author of which used the
Cartesian system of coordinates, and the presence of
crystal bending was taken into account by including an
additional potential energy into the Hamilton operator in the FW representation. In our notation, this
energy is
\begin{equation}
W=-\frac{p_\phi v_\phi x}{R},
\label{JEeq3} \end{equation} where $R$ is the crystal bending radius and $x=0$ corresponds to the middle of the distance between the crystal planes. During planar channeling in bent crystals,
the interplanar distance $d_p$ is negligibly small as compared with the crystal bending radius. Therefore, the
approach used in \cite{JETP1995} leads to the obtained correct
results. The author of this paper gave detailed quantum-mechanical description of the effects occurring
during particle and nucleus channeling in unbent and
bent crystals. For this reason, we restrict ourselves to
deriving well-known formulas \cite{BarJAPL,Lubo}, describing the
spin motion during planar channeling in bent crystals,
starting from the general equations obtained above.

At planar channeling, the field of planes is
characterized by the even potential $\Phi(x) = \Phi(-x)$. For
nuclei that move in the channeling mode and have
positive charges, this field can be approximated by the
harmonic potential
\begin{equation}
\Phi(x)=\frac{ax^2}{2}, ~~~ a=\frac{8U_0}{d_p^2},\label{JEgp}
\end{equation} where $U_0$ is the maximum value of the potential and
$d_p$ is the distance between the crystal planes. We assume
that the crystal is bent so that the bending plane is perpendicular to the crystal planes and the curvature
radius is $R$. In this case, the plane potential (\ref{JEgp}) becomes
\begin{equation}
\Phi(\rho)=\frac{a(\rho-R)^2}{2}. \label{JEgpl}
\end{equation} We do not consider the magnetic field and neglect
effects caused by possible electric dipole moments.

At channeling, the particle oscillates with
respect to the equilibrium trajectory which is a circular
arc. Therefore, the average force acting on the particle
in the cylindrical system of coordinates is zero. As follows from Eqs. (\ref{eqmff}) and (\ref{cllim}), in this case, in the classical limit, we have
$$e<\bm E>=<\bm\omega\times\bm\pi>.$$
Another form of this equation is
\begin{equation}
e<E_\rho>=-m|\omega|\sqrt{\gamma^2-1}. \label{JEgf}
\end{equation} The fact that the kinetic energy of transverse
motion is significantly smaller than that of longitudinal motion during channeling is taken into account in
Eq. (\ref{JEgf}).

Substitution into Eqs. (\ref{FWHamOme}) and (\ref{eq36JMP}) gives the angular velocity of the spin precession
\begin{equation} \bm\Omega=\frac1\gamma\left[\frac{g-2}{2}\left(\gamma^2-1\right)-1\right]\bm\omega. \label{FWOmevp} \end{equation}
When passing to the Cartesian system of coordinates,
the quantity $-\bm\omega$ is added to the angular precession
velocity, and we obtain the Lyuboshits formula \cite{Lubo} determining the relation between the angular velocities of the spin and momentum rotations:
\begin{equation}\begin{array}{c}
\bm\Omega^{(Car)}=\frac{\gamma-1}{\gamma}\left[\frac{g-2}{2}\left(\gamma+1\right)+1\right]\bm\omega.
\end{array} \label{JEnew} \end{equation}

The formula determining the effect of particle and
nucleus spin rotation during planar channeling in bent
crystals was first derived in a paper by V.G. Baryshevsky \cite{BarJAPL}.
The fast increase in the ratio $\Omega/\omega$ with increasing
Lorentz factor makes it possible to measure the magnetic moments of relativistic particles with short lifetimes \cite{BarJAPL}.

\section{Discussion and summary}

It is natural to choose the cylindrical system of
coordinates in the quantum-mechanical description
of relativistic spin 1/2 particles and nuclei channeled
in bent crystals. However, the problem of determining
projections of the spin operator on the radial and azimuthal directions appears in this case. Using ordinary
Dirac matrices for the given projections is its best solution, found in \cite{Schluter}. In this paper, the Dirac equation in
cylindrical coordinates obtained in \cite{Schluter} is supplemented with terms describing the AMM and EDM
(with strict substantiation of this procedure using
methods of quantum mechanics of Dirac particles in
gravitational fields). Although the form of the
obtained equation does not differ from that of the corresponding equation in Cartesian coordinates, the difference between them is manifested after transformation into the FW representation. In this representation, the Hamiltonian in cylindrical coordinates
differs from the corresponding Hamiltonian in Cartesian coordinates by the presence of an additional spin
term. Although the spin-independent parts of two
Hamiltonians determining the particle and nucleus
motion coincide formally, the obtained equations of
motion are significantly different. The equation of particle and nucleus motion in the cylindrical system
of coordinates includes the centrifugal force and also
determines the effect of the mutual influence of particle motion in the directions $\bm e_\rho$ and $\bm e_\phi$. This effect has a
kinematic nature and, in particular, is manifested in
the appearance of the oscillating force in the azimuthal direction during oscillatory motion in the
radial direction and vice versa. Passing to the classical
limit makes it possible to establish that the equations
of motion for the kinetic momentum and spin in the
two considered systems of coordinates agree completely with each other. As an example demonstrating
the correctness of the description of physical phenomena using the equations obtained in this paper, we have
derived a well-known relation between the angular
velocities of the spin and momentum vector rotations
during planar channeling in bent crystals.

Using methods developed in gravitation theory
makes it possible not only to provide the correct
description of particles and nuclei with the AMM and
the EDM in cylindrical coordinates, but also to determine the physical nature of the main specific features
of such a description. The analogy between the cylindrical system of coordinates and the rotating reference
system demonstrated previously in \cite{RPJSTAB} within the
framework of the classical approach appears paradoxical at first glance. Although the cylindrical system
of coordinates and the rotating reference system
belong to planar spacetime manifolds, the structures
of the metric tensor in these systems are significantly
different. The main physical properties of the rotating
reference system are determined by off-diagonal components of the metric tensor $g_{0i}$, and the metric is
stationary and nonstatic. As opposed to this, the
metric of the cylindrical system of coordinates (\ref{metrite}) is
static and has a single nontrivial component. However,
the calculation of gravitoelectromagnetic fields
explains the indicated analogy exhaustively. In both
cases, there is no gravitoelectric field, and the gravitomagnetic field $\bm{\mathcal{B}}$ is $\bm\omega u^{\widehat{0}}$ in the rotating reference system and $(u^{\widehat{\phi}}/\rho)\bm e_z$ in the cylindrical system of coordinates. If the condition $\bm\omega= \omega\bm e_z= [u^{\widehat{\phi}}/(\rho u^{\widehat{0}})]\bm e_z$ is satisfied, these quantities are equal to each other, and the
dynamics of particles and their spins in the two systems becomes identical. This condition can always be
satisfied for an individual particle, but it cannot be satisfied for an ensemble (beam) of particles with different momenta. The possibility of describing an ensemble of particles with an arbitrary distribution over
momenta is an important advantage of the cylindrical
system of coordinates.

Thus, in this paper, we presented the general quantum-mechanical description of relativistic spin 1/2
particles and nuclei channeled in bent crystals by using
the cylindrical system of coordinates. The results of
our study can be used to solve concrete problems of
channeling theory.

\vskip 3mm

I express my deep gratitude to V.G. Baryshevsky for
long-term collaboration and discussion of the
obtained results.

The work was supported by the Belarusian Republican Foundation for Fundamental Research.

\newpage


\begin{thebibliography}{}

\bibitem{Gang}
\emph{Gangrskii Yu.P.} // Soros. Obrazov. Zh. 2000. V. 6. No. 8. P.
93.

\bibitem{BarJAPL}
\emph{Baryshevsky V.G.} // Sov. Tech. Phys. Lett. 1979. V. 5. P. 73.

\bibitem{BarJPhys}
\emph{Baryshevsky V.G.} // J. Phys. G. 1993. V. 19. No. 2. P. 273.

\bibitem{Lubo}
\emph{Lyuboshits V.L.} // Sov. J. Nucl. Phys. 1980. V. 31. P. 509.

\bibitem{CAB}
\emph{Chen D., Albuquerque I.F., Baublis V.V. et al.} // Phys.
Rev. Lett. 1992. V. 69. No. 23. P. 3286.

\bibitem{KSCC}
\emph{Khanzadeev A.V., Samsonov V.M., Carrigan R.A. and Chen D.}
// Nucl. Instr. Meth. B. 1996. V. 119. No. 5. P. 266.

\bibitem{JETP1995}
\emph{Silenko A.J.} // J. Exp. Theor. Phys. 1995. V. 80. P. 690. 

\bibitem{Schluter}
\emph{Schl\"{u}ter P., Wietschorke K.-H. and Greiner W.} 
// J. Phys. A: Math. Gen. 1983. V. 16. No. 9. P. 1999.

\bibitem{JMP}
\emph{Silenko A.J.} // J. Math. Phys. 2003. V. 44. Iss. 7. P. 2952.
%

\bibitem{RPJ} \emph{Silenko A.J.} // Russ. Phys. J. 2005. V. 48. P. 788.

\bibitem{PauliFT}
\emph{Pauli W.} // Ann. Inst. Henri Poincar\'{e}. 1936. V. 6. No. 2. P. 109.

\bibitem{HN}
\emph{Hehl F.W. and Ni W.T.} // Phys. Rev. D. 1990. V. 42. Iss. 6. P. 2045.

\bibitem{BlagojevicHehl}
\emph{Blagojevi\'c M. and Hehl F.W.} (eds.) Gauge Theories of Gravitation.
A Reader with Commentaries. London: Imperial College Press, 2013. 635 p.

\bibitem{LL2}
\emph{Landau L.D. and Lifshitz E.M.} Course of Theoretical
Physics, Vol. 2: The Classical Theory of Fields (Pergamon, Oxford, 1975).

\bibitem{ostrong}
\emph{Obukhov Yu.N., Silenko A.J., and Teryaev O.V.} //
Phys. Rev. D. 2011. V. 84. Iss. 2. P. 024025.

\bibitem{OSTgrav} \emph{Obukhov Yu.N., Silenko A.J., and Teryaev O.V.} // 
Phys. Rev. D. 2013. V. 88. Iss. 8. P. 084014.

\bibitem{PK}
\emph{Pomeranskii A.A. and Khriplovich I.B.} //
J. Exp. Theor.
Phys. 1998. V. 86. P. 839.

\bibitem{Warszawa}
\emph{Silenko A.J.} //
Acta Phys. Polon. B Proc. Suppl. 2008. V. 1. No. 1. P. 87. 

\bibitem{OST}
\emph{Obukhov Yu.N., Silenko A.J., and Teryaev O.V.} //
Phys. Rev. D. 2009. V. 80. Iss. 6. P. 064044.

\bibitem{RPJSTAB}
\emph{Silenko A.J.} // Phys. Rev. ST Accel. Beams. 2006. V. 9. Iss. 3. P. 034003.

\bibitem{PRA}
\emph{Silenko A.J.} // 
Phys. Rev. A. 2008. V. 77. Iss. 1. P. 012116.

\bibitem{JINRLClL}
\emph{Silenko A.J.} // Phys. Part. Nucl. Lett. 2013. V. 10. No. 2. P. 91.

\bibitem{TMPFW}
\emph{Silenko A.J.} // 
Theor. Math. Phys. 2013. V. 176. No. 2. P. 987. 

\bibitem{PRD2}
\emph{Silenko A.J. and Teryaev O.V.} // 
Phys. Rev. D. 2007. V. 76. № 6. P. 061101(R).

\end{thebibliography}
\end{document}